\documentclass[12pt]{article}
\usepackage{a4wide}
\usepackage{amsfonts}
\usepackage{amssymb}
\newcommand{\ns}{\ \indent} 
\newcommand{\Ima}{\mbox{Im}}
\newcommand{\sigmabf}{\mbox{\boldmath $\sigma$}}
\newcommand{\R}{\bf R}
\begin{document}
\title{\vskip -70pt
  \begin{flushright}
    {\normalsize{DAMTP-1999-52, hep-th/9905009}} 
  \end{flushright}
  \vskip 15pt
  {\bf \Large \bf Instanton vibrations of the 3-Skyrmion}
  \vskip 10pt}
\author{ 
  Conor J. Houghton,\thanks{Email: C.J.Houghton@damtp.cam.ac.uk}
\\[5pt] 
{\normalsize {\sl Department of Applied Mathematics
      and Theoretical Physics,}}\\
  {\normalsize {\sl University of Cambridge, Silver Street,
      Cambridge, CB3 9EW, United Kingdom.}}\\[10pt]
}

\date{April 1999}
 
\maketitle

\begin{abstract}
  
  The Atiyah-Drinfeld-Hitchin-Manin matrix corresponding to a
  tetrahedrally symmetric 3-instanton is calculated. Some small
  variations of the matrix correspond to vibrations of the
  instanton-generated 3-Skyrmion. These vibrations are decomposed
  under tetrahedral symmetry and this decomposition is compared to
  previous knowledge of the 3-Skyrmion vibration spectrum.
\end{abstract}

\section{Introduction}

\ns In the Skyrme model, the classical $B$-nucleon nucleus is a
$B$-Skyrmion: a minimum energy Skyrme field with topological charge
$B$. The $B$-Skyrmions have been calculated numerically for $B$ up to
nine \cite{BTC,BS}.

The Skyrme model is nonrenormalizable and so cannot be quantized
as a field theory. However, it is hoped that the quantum mechanics on
some finite-dimensional space in the charge $B$ sector might give a
good model of the quantized $B$-nucleon. This approach has been
reasonably successfully in the 1-Skyrmion case \cite{ANW} but for
higher $B$ it is hard to choose a suitable, tractible, finite-dimensional
space.  

The 1-Skyrmion is spherically symmetric and has six zero modes: three
translational and three rotational.  This suggests that the
finite-dimensional space should be $6B$-dimensional and one popular
candidate is the gradient-flow manifold descending from the charge $B$
spherical saddle-point \cite{M}.  Recently, the vibration spectra of
$B$-Skyrmions have been calculated numerically for $B$ equals two,
three, four and seven \cite{BBT,B}. It was found that the the
vibration frequencies of the $B$-Skyrmion are divided into two groups
by the breather mode which corresponds to dilation.  This suggests
that it might be necessary to add the breather mode to the
$6B$-dimensional space to give $(6B+1)$-dimensions, or even to include
seven dimensions for each Skyrmion to give a $7B$-dimensional space.
Another suggestion is that $8B-3$ vibrational modes should be expected
\cite{BM}.

The modes below the breather have been interpreted as being
monopole-like and may correspond to the gradient-flow manifold
descending from the saddle-points of infinite Skyrmion separation and
from the charge $B$ torus \cite{HMS}. It may be that this is the space
upon which the quantization should be performed. It is a
$(4B+2)$-dimensional space.

All these spaces are thought to be well approximated by instanton
generated Skyrme fields \cite{AM1,AM2}. In the instanton construction,
Skyrme fields are derived from instanton fields by calculating their
holonomy in the $x_4$ direction \cite{AM1}. There is a
$(8B-1)$-dimensional family of baryon number $B$ Skyrme fields derived
from the space of $B$-instantons. It is known for $B$ equals one, two,
three and four that the $B$-Skyrmion is well approximated by an
instanton-generated Skyrme field \cite{AM1,LM}. In this paper the
vibrations around the instanton generated 3-Skyrmion are studied. The
decomposition of these vibrations as representations of the
tetrahedral group includes the same representations as are found in
the decomposition of the numerically determined spectrum. This seems
to indicate that the numerically determined vibrations are close to
being tangent to the space of instanton generated Skyrme fields. It is
consistient with the view that, whatever space should be used to
quantize the $B$-Skyrmion, it is approximated by a subspace of the
space of instanton generated Skyrme fields.

The 3-Skyrmion has tetrahedral symmetry \cite{BTC}. In \cite{LM}, the
Jackiw-Nohl-Rebbi (JNR) ansatz \cite{JNR} is used to derive a
tetrahedral 3-instanton. From this, the instanton-generated
3-Skyr\-mion is calculated.  In \cite{W}, Walet examines vibration
modes of the instanton-generated 3-Skyrmion by varying the JNR
parameters.  Although a large class of instantons can be constructed
using the JNR ansatz, it is not general. However, the
Atiyah-Drinfeld-Hitchin-Manin (ADHM) construction \cite{ADHM,CFTG,CWS} is
general and, in this paper, the tetrahedral 3-instanton ADHM matrix
is calculated. The instanton-generated 3-Skyrmion vibration modes are
then examined by varying the ADHM parameters. The vibration
frequencies are not calculated. However, the vibrations are decomposed
under the action of the tetrahedral symmetry. This allows the
decomposition to be compared to other calculations of 3-Skyrmion
vibrations.

\section{The ADHM matrix for the 3-Skyrmion}

\ns Symmetric ADHM matrices have been discussed in a recent paper
by Singer and Sutcliffe \cite{SS} and this should be consulted
for any details not included in this section. 

The ADHM matrix for a $B$-instanton is a quaternionic matrix
\begin{equation}
\hat{M}=\left(\begin{array}{c}L\\M\end{array}\right)
\end{equation}
where $L$ is a $B$-vector and $M$ is a symmetric $B\times B$
matrix. $\hat{M}$ must satisfy the ADHM constraint
\begin{equation}
\hat{M}^\dagger \hat{M}\mbox{ is real}.\label{ADHMc} 
\end{equation}
Dagger denotes quaternionic conjugation and matrix transposition.

Pure quaternions can be identified with $\mathfrak{su}_2$ by
$-i\sigmabf=(i,j,k)$ where $\sigmabf=(\sigma_1,\sigma_2,\sigma_3)$ are
the Pauli matrices. With this identification, the instanton gauge
fields are
\begin{equation}
A_\mu(x)=N^\dagger(x) \partial_\mu N(x)
\end{equation}
where $N(x)$ is the unit length $(B+1)$-vector solving 
\begin{equation}\label{coneq}
N^\dagger(x) \left(\begin{array}{c}L\\M-x{\bf 1}_B\end{array}\right) =0.
\end{equation}
In this equation, ${\bf 1}_B$ is the $B\times B$ identity matrix and
the $\R^4$ position is written as a quaternion: $x=x_4+x_1i+x_2j+x_3k$.

There is an ambiguity in choosing $N(x)$ given by 
\begin{equation}
N(x)\rightarrow N(x)g(x)
\end{equation}
where $g(x)$ is a unit quaternion. The unit quaternions are identified
with the two-dimen\-sional representation of SU$_2$ and so this ambiguity
corresponds to gauge transformations of the fields. There is also an
ambiguity in $\hat{M}$ given by
\begin{equation}\label{Mhatamb}
\hat{M}\rightarrow\left(\begin{array}{cc} g&0\\0& \rho \end{array}\right) \hat{M} \rho^{-1}
\end{equation}
where $g$ is a unit quaternion and $\rho$ is a real orthogonal
$B\times B$ matrix. This is a gauge transformation of the ADHM matrix:
it does not affect the fields.

This convenient version of the ADHM data is the canonical form
discussed in \cite{CFTG,CWS}. The ADHM construction, as originally
introduced, involved a larger gauge ambiguity and a second ADHM
matrix: a matrix coefficient of $x$ in (\ref{coneq}). The canonical
form is a partial fixing of the larger gauge ambiguity.

Under the conjugate action of unit quaternions on $x$, the real part
of $x$ is fixed and the imaginary part transforms under the
three-dimensional representation of SO$_3$. This means that for a
spatial rotation $R$ there is a quaternion $g$ so that
\begin{equation}
x_4+(i\,j\,k)R\left(\begin{array}{c}x_1\\x_2\\x_3\end{array}\right)=gxg^{-1}.
\end{equation}
Of course, $-g$ corresponds to the same $R$: SU$_2$ is a double cover
of SO$_3$.

As explained in \cite{SS}, an instanton has the spatial rotation
symmetry $x\rightarrow g x g^{-1}$ for unit quaternion $g$ if
\begin{equation}
 \left(\begin{array}{c}L\\M-gxg^{-1}{\bf 1}_B\end{array}\right)
=\left(\begin{array}{cc} \tilde{g}&0\\0& \rho\cdot g \end{array}\right)
  \left(\begin{array}{c}L\\M-x{\bf 1}_B\end{array}\right) g^{-1}\cdot
  \rho^{-1}
\label{invcond}
\end{equation}
where $\tilde{g}$ is a unit quaternion and $\rho\cdot g$ is the
product of the real orthogonal matrix $\rho$ and the unit quaternion
$g$. Thus, the ADHM matrix is symmetric if the spatial rotation is
equivalent to a gauge transformation. If the instanton is symmetric
under some subgroup of SO$_3$ then the collection of $\rho$'s and of
$\tilde{g}$'s form a real $B$-dimensional representation and a complex
two-dimensional representation of the corresponding binary subgroup of
SU$_2$.

\subsection{The tetrahedrally symmetric ADHM matrix}

\ns Since the 3-Skyrmion is tetrahedrally symmetric, the corresponding ADHM
matrix is also tetrahedrally symmetric. The tetrahedral group $T$ is the twelve
element subgroup of SO$_3$ which, in one orientation, is generated by a
rotation of $\pi$ about the $x_3$-axis and a rotation of $2\pi/3$
about $x_1=x_2=x_3$.  These generators will be called $r$ and $t$
respectively and the corresponding unit quaternions are $g(r)=k$ and 
$g(t)=(1-i-j-k)/2$.

The group is isomorphic to the alternating group ${\mathfrak A}_4$.
The tetrahedral double group is the 24 element subgroup of SU$_2$
which double covers the tetrahedral group. The representation theory
of the tetrahedral group is described in, for example, Hamermesh
\cite{H}. There are one, two and three dimensional representations
derived by restricting the one, two and three dimensional irreducible
representations of SU$_2$ to the tetrahedral group. They are
$A={\underline 1}|_T$, $E'={\underline 2}|_T$ and $F={\underline
  3}|_T$ where ${\underline n}$ denotes the irreducible
$n$-dimensional representation of SU$_2$. There is, in addition, the
two-dimensional representation $E$ and the four-dimensional
representation $G'$. These representations are reducible
into conjugate pairs of representations with complex characters.

The ADHM matrix
\begin{equation}
\hat{M}_T=\left(\begin{array}{c}L_T\\M_T\end{array}\right)=\left(\begin{array}{ccc}i&j&k\\0&k&j\\k&0&i\\j&i&0\end{array}\right)
\end{equation}
is tetrahedrally symmetric. This matrix was found by trial and error.
Having written down a likely form of the matrix it is easy to check whether
or not it has the required symmetries. Explicitly, the matrices giving
the compensating gauge transformations are
\begin{equation}
\rho(r)=\left(\begin{array}{ccc}-1&0&0\\0&-1&0\\0&0&1\end{array}\right)
\end{equation}
with $\tilde{g}(r)=g(r)$ for $r$ and
\begin{equation}
\rho(t)=\left(\begin{array}{ccc}0&1&0\\0&0&1\\1&0&0\end{array}\right)
\end{equation}
with $\tilde{g}(t)=g(t)$ for $t$.  Thus, in this case the $\rho$'s form the
representation $F$ and the
$\tilde{g}$'s form the representation $E'$.  

$\hat{M}_T$ is not just symmetric under $T$. It is also symmetric under the 24
element group $T_d$ which extends $T$ by the $S_4$ element $u$:
\begin{equation}
u:(x_1,x_2,x_3,x_4)\rightarrow (-x_2,x_1,-x_3,-x_4).
\end{equation}
In fact,
\begin{equation}
\left(\begin{array}{c}L\\M-u(x){\bf 1}_B\end{array}\right)
=-\left(\begin{array}{cccc}-1& \\&\rho\end{array}\right)\frac{1-k}{\sqrt{2}}\left(\begin{array}{c}L\\M-x{\bf 1}_3\end{array}\right)\frac{1+k}{\sqrt{2}}\,\rho^{-1}
\end{equation}
where 
\begin{equation}
\rho=\left(\begin{array}{ccc}0&1&0\\-1&0&0\\0&0&1\end{array}\right).
\end{equation}
The representation theory for $T_d$ is also described in Hamermesh
\cite{H}. The vector representation $F$ of $T$ is replaced by a true
vector $F_2$ and an axial vector $F_1$. In the same way, the trivial
representation $A$ is replaced by a true scalar $A_1$ and a
pseudo-scalar $A_2$. Although reducible as a
representation of $T$, $E$ is irreducible as a representation of
$T_d$. There are similar changes to the double group representations.

\subsection{Uniqueness and the tetrahedrally symmetric ADHM matrix}

\ns $\hat{M}_T$ is not unique, there is a two-parameter family of
tetrahedral matrices given by $x(\hat{M}_T+y{\bf 1}_4)$. $y$ can be set to
zero. It corresponds to translation of the instanton in the $x_4$
direction and this does not change the corresponding Skyrme field. $x$
is a scale parameter and, when calculating the instanton-generated
Skyrmion, the scale is fixed by minimization of the Skyrme energy.
There can be no more than two parameters because, as explained in
\cite{SS}, it follows from (\ref{invcond}) that
\begin{equation}
M_T:\underline{3}|_T\rightarrow (\underline{3}\times\underline{2}\times\underline{2})|_T
\end{equation}
that is 
\begin{equation}
M_T:F\rightarrow 3F+E+A
\end{equation}
and so there is a three-parameter family of candidate $M_T$ matrices. This
is exhausted by
\begin{equation}
\left(\begin{array}{ccc}d&qk&rj\\rk&d&qi\\qj&ri&d\end{array}\right).
\end{equation}
Similarly,
\begin{equation}
L_T:(\underline{3}\times\underline{2})|_T\rightarrow \underline{2}|_T=E'.
\end{equation}
and, since $(\underline{3}\times\underline{2})|_T=E'+G'$, there is a
one-parameter family of $L_T$.  The symmetry of $M_T$ and the ADHM
constraint (\ref{ADHMc}) reduce these four parameters to the two
parameters $x$ and $y$ above. There is another two-parameter family of
symmetric matrices corresponding to the dual tetrahedron. This is
given by replacing $M_T$ in $\hat{M}_T$ by $-M_T$. 

It is possible to translate between JNR data and ADHM matrices. This
is useful here because it gives an explicit verification that the
3-Skyrmion generating instanton of \cite{LM} lies in the one-parameter
family $x\hat{M}_T$. The general formula, translating JNR data into
ADHM data, is given in Section 5 of \cite{CFTG}. Unfortunately, these
ADHM data are not in the canonical form involving a single ADHM
matrix. It seems that it is difficult to write the general JNR-derived
ADHM data in canonical form. However, in the particular case of
interest here, a straight-forward calculation shows that $\hat{M}_T$
is the canonical form of ADHM data derived from tetrahedral JNR data. 

\section{Variations}

\ns Small variations around $\hat{M}_T$ are now considered. Writing
\begin{equation}
\hat{M}=\hat{M}_T+\hat{m}
\end{equation}
$\hat{m}$ satisfies the linearized ADHM constraint
\begin{equation}
\Ima(\hat{m}^\dagger \hat{M}_T+\hat{M}_T^\dagger\hat{m})=0.
\end{equation}
If
\begin{equation}
\hat{m}=\left(\begin{array}{ccc}l_1&l_2&l_3\\m_{11}&m_{12}&m_{13}\\m_{12}&m_{22}&m_{23}\\m_{13}&m_{23}&m_{33}\end{array}\right)
\end{equation}
the linearized equations are
\begin{eqnarray}
\Ima(\bar{l}_1j+\bar{m}_{11}k+\bar{m}_{13}i-il_2-km_{22}-jm_{23})&=&0,\\
\Ima(\bar{l}_1k+\bar{m}_{11}j+\bar{m}_{12}i-il_3-km_{23}-jm_{33})&=&0,\nonumber\\
\Ima(\bar{l}_2k+\bar{m}_{22}i+\bar{m}_{12}j-jl_3-km_{13}-im_{33})&=&0\nonumber
\end{eqnarray}
where bar denotes quaternionic conjugation. These equations can be
solved to give expressions for nine of the twelve $l_i$
parameters. The remaining three components correspond to the gauge
transformation
\begin{equation}
\hat{M}_T\rightarrow\left(\begin{array}{cc} g&0\\0& {\bf 1}_3 \end{array}\right) \hat{M}_T.
\end{equation}
This gauge freedom can be fixed by requiring, for example, that
$l_1$ is proportional to $i$. In this way,  $l$ may be
completely determined by $m$ and by gauge fixing. 

In order to decompose the 24 $m_{ij}$ component as representations of
$T_d$, the actions of $r$, $t$ and $u$ on $\hat{M}_T$ are considered.
Thus, for example,
\begin{equation}
\rho(r)g(r)mg(r)^{-1}\rho(r)^{-1}=\left(\begin{array}{ccc}-km_{11}k&-km_{12}k&km_{13}k\\-km_{12}k&-km_{22}k&km_{23}k\\km_{13}k&km_{23}k&-km_{33}k\end{array}\right)
\end{equation}
and so the character of $r$ is zero. The characters of $t$ and $u$ can
be calculated in the same way: they are also zero. This means that the
decomposition of the $m_{ij}$ is
\begin{equation}
A_1+A_2+2E+3F_1+3F_2.
\end{equation} 

Not all of these multiplets correspond to Skyrmion vibrations. There
remains the gauge freedom
\begin{equation}
M_T\rightarrow \rho M_T \rho^{-1}.
\end{equation}
By considering an infinitesimal $\rho$ and calculating the character,
it is found that this variation is an $F_1$. The remaining 21
variations correspond to the 21 dimensions of the space of
3-instantons. However, the variation corresponding to time
translation, $A_2$, does not affect the 3-Skyrmion and is discarded.
Twenty variational modes remain: six of these correspond to zero
modes. In fact, the translation and rotation zero modes of the
Skyrmion correspond to an $F_2$ and an $F_1$ respectively. Thus, the
instanton modes of the 3-Skyrmion decompose as
\begin{equation}
A_1+2E+2F_1+3F_2 \label{idecomp}
\end{equation}
under $T_d$ and, of these, one $F_1$ and one $F_2$ are zero modes and
the rest are vibrational modes. The isospin zero modes are not
included in this decomposition because they do not correspond to
variations of the ADHM matrix.

\section{Discussion}

\ns The $A_1$ is the breather mode corresponding to dilation. In the
numerical results of Baskerville, Barnes and Turok \cite{BBT}, it
appears in the middle of the vibration spectrum.  In order of increasing
frequency and ignoring radiation, Baskerville, Barnes and Turok find
the spectrum to be 
\begin{equation}
F_2+E+A_1+F_2+E.
\end{equation}
The $E+F_2$ below the breather are the modes described in \cite{HMS}
as monopole modes. They correspond to variations of the rational map
parameters in the rational map ansatz of \cite{HMS}.  

The $E+F_2$ above the breather are discussed in \cite{BM} by
Baskerville and Michaels. It has been observed that, to a good
approximation, there are $2B-2$ straight lines of zero baryon density,
known as branch lines, radiating from the centre of a $B$-Skyrmion.
In \cite{BM}, the variations of the angular positions of the branch
lines are parametrized. These are then decomposed.  It is noted that
if an axial vector is removed from this decomposition, the
decomposition then matches the super-breather modes in the 2-Skyrmion
and 3-Skyrmion spectra. In the 4-Skyrmion case the decomposition is
consistent with the observed spectrum. Baskerville and Michaels
interprete the axial vector which must be removed as the axial vector of
rotational zero modes.

Some of the monopole modes also change the branch line positions.
Therefore, \cite{BM} implies that the super-breather mode
decomposition duplicates part of the decomposition of the monopole
modes.  There are, in total, $4B+2$ monopole modes of a $B$-Skyrmion.
Six of them do not change the branch lines: of these, three are the
isospin zero modes. Because the parameters in the rational map are
complex, the monopole modes come in pairs of opposite parity.  There
are three monopole modes which compose pairs with isospin modes in
this way. These are the other three modes which do not change the
positions of branch lines. The remaining $4B-4$ modes change the
positions of the branch lines.  These $4B-4$ modes include the three
translational and three rotational zero modes along with $4B-10$
vibrational modes.  Thus, there are $4B-7$ monopole modes which
are not rotational zero modes and which change the positions of the
branch lines. It is possible to reformulate the observation made in
\cite{BM}: the decomposition of these $4B-7$ modes
duplicates the decomposition of the super-breather modes. 

Thus, an exception is made for the rotational zero modes, the translational
zero modes are duplicated but the rotational zero modes are not. In
fact, in the $B=3$ case, the instanton modes contain a $F_1$
duplicating the rotational zero modes. This $F_1$ has not been
observed numerically.  The reason for this may be that the $F_1$ mode
has a rather high frequency.

For $B=3$, the monopole modes which fix the branch line positions are
an $F_1$ of isospin and an $F_2$ vibration. The monopole modes which
change the branch line positions are the rotational and translational
zero modes $F_1+F_2$ and the multiplet of vibration modes $E$.  Thus,
the $E$ in the super-breather part of the 3-Skyrmion spectrum
duplicates the $E$ in the monopole part. The $F_2$ duplicates the
translational zero modes. If a duplicate is also included for the
rotational zero modes; then the aggregate of the breather, the
monopole modes and their duplicates matches the instanton mode
decomposition (\ref{idecomp}).

Because Walet uses JNR ansatz instantons, not all of the instanton modes
are included in the harmonic analysis of \cite{W}. To be precise, there
is only one $E$, whereas the ADHM construction gives two. In the case
of the rational map decomposition, it is known that the $E$ is spanned
by tangent vectors lying along the $S_4$ symmetric geodesics. These are referred
to in \cite{HS} as twisted line scattering geodesics. 

There is a three-dimensional family of ADHM matrices, symmetric under
the $D_{2d}$ generated by $u$ and $\pi$ rotations about the Cartesian
axes. Vibrations tangent to this family lie in the $A_1+2E$ of the
decomposition (\ref{idecomp}).  The symmetric ADHM matrices are
\begin{equation}
\hat{M}_{D_{2d}}=\left(\begin{array}{ccc}ai&aj&bk\\e&ck&dj\\ck&-e&di\\dj&di&0\end{array}\right)
\end{equation}
with, from the ADHM constraint,
\begin{eqnarray}
ab+de-dc&=&0\\
d^2-a^2+2ec&=&0.\nonumber
\end{eqnarray}
For $a=b=c=d=x$ and $e=0$ this is $\hat{M}_T$ and for $a=c=d=e=0$ it
is axially symmetric about the $x_3$-axis.  Translating $D_{2d}$ JNR
data to ADHM data and rewriting it in canonical form gives the
two-parameter subfamily with $a=d$, $c=b$ and $e=0$. However, as noted
by Walet, this subfamily of $D_{2d}$ symmetric ADHM matrices shares the
curious feature of the $T_h$ matrices in \cite{SS}, it does not
include well-separated instantons of equal scale. More complicated
subfamilies, including well-separated instantons of equal scale, may
be chosen by using arguments similar to those in \cite{SS}. One simple
example, with $b=1$ fixing the scale, is
\begin{equation}
e=\frac{c(c^2-1)}{2(c^4+1)}.
\end{equation}
Of course, this is just one path which passes though the various
features associated with twisted line scattering. The infinitesimal
behaviour around $\hat{M}_T$ does not give the splitting of the $2E$
into a sub-breather $E$ and the super-breather $E$. It is not known
how to make this split, without calculating the holonomy and
performing the full harmonic analysis as Walet did for JNR ansatz
instanton-generated 3-Skyrmions.

In conclusion, the instanton modes of the 3-Skyrmion have been
calculated and decomposed. The decomposition fits well with other
similar decompositions. The primary question provoked by the
calculations is whether it is possible to split the modes
further without undertaking the harmonic analysis.

\section*{Acknowledgments}
\ns The financial assistance of Fitzwilliam College, Cambridge is
gratefully acknowledged. This work is supported, in part, by PPARC. I
am grateful to Kim Baskerville for useful discussion.

\end{document}